# A VR-based Priming Framework and Technology Implementation to Improve Learning Mindsets and Academic Performance in Post-Secondary Students

Dan Hawes and Ali Arya

**Abstract**

Recent research indicates that most post-secondary students in North America "felt overwhelming anxiety" in the past few years, negatively affecting well-being and academic performance. Further research revealed that other emotions, biases, perceptions, and negative thoughts, can similarly affect student academic performance. To address this problem, we classify these counterproductive mindsets, including anxiety, into *Scarcity Mindset,* a self-limiting perspective that appropriates cognitive bandwidth required for essential processes like learning in favour of addressing more critical needs or perceived insufficiencies. Through a multi-disciplinary literature analysis of ideas in cognitive science, learning theories and mindsets, and current technology approaches that are suited to address the limitations of scarcity thinking, we identify strategies to help transition students to a more positive *Abundance Mindsets*. We demonstrate that these priming intervention strategies can transfer to leading edge digital environments, particularly Virtual Reality (VR). Offering further insights into the findings of our two previously presented studies, we argue that priming interventions related to preparatory activities and the context priming are transferable to virtual reality environments. As such, building on our multidisciplinary research insights, we propose a comprehensive priming model that exploits priming techniques in an iterative process called Cyclical Priming Methodology (CPM). These intervention strategies can focus on student preparation, motivation, reflection, the context of the learning environment, and other aspects of the learning process. Building on CPM, we further propose a technology implementation within VR called Virtual Reality Experience Priming (VREP) and discuss the process to embed CPM/VREP activities within the Experiential Learning Theory (ELT) cycle.

**Keywords:** Priming, Scarcity, Virtual Reality, Situated Learning

## 1. Introduction

Recent statistics indicate that student anxiety has reached concerning levels in post-secondary institutions affecting students' mental health and academic performance. A 2018 survey by the American/National College Health Association observed that over 62% of students felt overwhelming anxiety within the past year (ACHA/NCHA II, 2018), a figure which was reported to worsen later, especially due to COVID-19 pandemic (Grubic et al., 2020; Broglia et al., 2021). While this data was alarming, anxiety is not the only counterproductive thought process affecting student academic performance. Further research and interdisciplinary literature reviews have revealed that other emotional states (negative learning mindsets, cognitive biases, stereotype perceptions, and poverty) also affect cognitive student academic performance (Arya et al., 2019; Banakou et al., 2018; Chang et al., 2019). A 2013 study demonstrated how seeding the idea of economic difficulty in certain participant groups (by exposing them to a fictional financial problem), had an immediate and profound negative impact on cognitive performance of certain (poorer) participants (Mullainathan and Shafir, 2013). The researchers attributed this effect to the induction of *scarcity mindset*: a limiting perspective that appropriates cognitive capacity required for processes like learning to focus on critical needs or perceived insufficiencies. Further research into cognitive science, learning theories, and motivational mindsets substantiated the negative impact of counterproductive mental states (Ciani et al., 2010; Dweck, 2008; Mullainathan and Shafir, 2013), but similarly, also offered possibilities and potential antidotes to scarcity thinking like the use of *priming* interventions to induce positive affect (Isen et al., 1987; Ashby and Isen, 1999), to change environments and imagery (Bhagwatwar et al., 2013; Burke et al., 2014), and to improve mental mindsets for learning (Yeager et al., 2019; Paunesku et al., 2015; Dweck, 2008).

Priming is generally defined as using a stimulus or action to affect the information processing availability for another idea or experience (Dennis et al., 2012). Behaviourism and conditioning theories provide key insights on positive affect inducement, repeated exposure techniques, and how to develop priming activities through associative learning and habituation (Wagner, 1977; Zajonc, 1968; Carl et al., 2019). These and others, such as experiential learning theory (ELT) (Kolb, 1984), provide insights into what methods of priming should be considered and when, during the learning cycle. Emerging technologies like Virtual Reality (VR), Video Games (VG), Artificial Intelligence (AI), and biometrics offer unique affordances that can contribute to improving the learning experience. For example, casual video games (CVG) may allow students to escape into a state of flow as an alternative to anti-depressants (Fish et al., 2018; Csikszentmihalyi, 2000). VR technology can be valuable in therapeutic contexts like cognitive-based training therapy, Virtual Reality Exposure Therapy (VRET) (Carl et al., 2019). VR has also been shown to have the potential to improve learning outcomes by offering more creative situated learning environments (Hawes and Arya, 2022a; Bhagwatwar et al., 2013).

Hence, our research is motivated by the problem of the counterproductive thought processes that negatively impact learning. We used the concept of scarcity mindset to group these thought processes and specifically addressed the problem of using technology, particularly Virtual Reality (VR) to offer priming interventions that ameliorate the negative impact of scarcity mindset on student academic performance. VR offers affordances such as customizable 3D simulation, immersion, empathy, presence, and embodiment (Shin, 2017), which make it a good candidate for many educational scenarios. These affordances, together with the increased use of VR for educational purpose, especially during the COVID-19 pandemic (Deponti et al., 2020; Singh et al., 2020, Sampaio et al., 2021) motivated us to focus on VR as a platform to provide educational priming. While priming is generally defined as a stimulus or action to affect the information processing availability for a subsequent idea or experience, we define priming in a learning context, to include any deliberate activity or stimuli intended to improve the student's mindset.

Various priming techniques have been shown to improve mental mindsets and learning outcomes. There is also research evidence such as our own previous work (Hawes and Arya, 2021; Hawes and Arya, 2022a; Hawes and Arya, 2022b) that suggest these priming techniques can be transferred to digital (VR) realm. Previously, and as presented in two separate conferences (Hawes and Arya, 2021; Hawes and Arya, 2022a), we had completed two independent research studies focused on the transferability and effectiveness of two priming methods: priming prior to the learning activity, classified as Preparatory Experience Priming (PEP), and priming through custom-designed, situated learning context, referred to as Context-Oriented Priming (COP). These two types were chosen as they are most commonly used in priming literature (Brown et al., 1989; Lave and Wenger, 1991; Bhagwatwar et al., 2013). Our first study focused on pre-learning (preparatory) priming through games and meditation, based on their success in non-VR environments (Mrazek et al., 2014; Fish et al., 2018). It showed that priming methods work and are transferable to VR, to reduce scarcity mindset and increase cognitive availability. The second study sought to extend our understanding of priming strategies to include context-oriented priming throughout the learning experience. We compared three different contexts for an animation course and demonstrated the effectiveness of VR-based context priming.

However, there is very limited theoretical work on how to organize priming interventions throughout learning process, what types of priming interventions to use, and how to implement them using digital technologies, especially VR. In this paper, we try to address these shortcomings through a multi-disciplinary literature review to establish a theoretical framework, critically reviewing our own past studies on the use of VR for educational priming to offer new insights on what could be learned from the findings, and finally proposing a new theoretical model called Cyclical Priming Methodology (CPM) and its VR implementation, Virtual Reality Experience Priming (VREP). We also discuss the process of embedding CPM/VREP activities within the Experiential Learning Theory (ELT) cycle (Kolb, 1984).

As a main contribution, this paper proposes a novel educational priming framework that intervenes strategically within the experiential learning cycle process. While the intent of the CPM is to transform scarcity thinking into abundance thinking in a real-time manner and further exploit the affordances of VR with our virtual reality experience priming (VREP) implementation, strategic interventions within the ELT cycle adds the critical element of timing. The preparatory experience priming (PEP) activity precedes the concrete experience and serves to increase cognitive bandwidth prior to the academic activity while within the context of the experience (Hawes and Arya, 2021). Situated learning environments may also induce context-oriented priming (COP) effects further contributing to the increase in cognitive bandwidth and overall academic performance (Hawes and Arya, 2022a). Based on the validation of PEP and COP, and further reflection, we propose other examples of additional timed interventions within the ELT cycle: motivation oriented prime (MOP), reflective experience prime (REP), direct attentional prime (DAP), visual imagination prime (VIP), creative improvisational prime (CIP), and prosocial inclusion prime (PIP) initiated at various stages within the ELT cycle.

## 2. Related Work

### 2.1 Anxiety, Biases, and Scarcity Mindset

Anxiety can have an extremely negative impact on learning performance and general health (Beiter et al., 2015) (Russell and Topham, 2012). Cue Utilization Theory (Easterbrook, 1959) suggests a narrowing of focus that restricts our utilization of environmental cues; cues that may be very relevant to our learning situations. The weapon focus effect (Loftus et al., 1987) is perhaps the most extreme but best example of what stress and anxiety can do to performance. When viewing a situation where a gun is present (e.g., a holdup at a store), to deal with the immediate threat, a process of cognitive tunneling is induced that excludes other relevant information in the environment.

While the negative effects of chronic anxiety on cognitive performance are generally accepted, the research suggests that anxiety is not the only counterproductive thought process affecting student academic performance. Further research reveals that other emotional states (negative mindsets, biases, stereotype perceptions, and poverty) also affect student academic performance (Arya et al., 2019; Banakou et al., 2018; Chang et al., 2019), not just anxiety. For example, a 2013 study demonstrated how seeding the idea of economic scarcity with poorer participant groups, without their conscious awareness, had an immediate and profoundly negative impact on academic performance (Mullainathan and Shafir, 2013).

Cognitive biases are lenses through which we experience our world. Fear/Negativity bias for example, genetically wires us to focus disproportionately on worst case scenarios that may keep us safe. Assuming a movement in the bush ahead is a lion rather than a wind disturbance may save our life, while the cost of assuming the worst-case scenario is generally negligible considering the potential risk (Ito et al., 1998; Soroka et al. 2018; Mathews and Macleod, 2002). Implicit bias, stereotypes, and prejudices, precondition us to analyse, categorize, and favour the familiar (Greenwald and Krieger, 2006; Marks and Nesse, 1994). This genetic predisposition can be effective in some situations but can easily run amok as the context of modern life changes.

Attentional Bias refers to the concept of hyperattention to threatening material where we selectively attend to a stimulus in the environment while tending to overlook or ignore other relevant stimuli (MacLeod et al., 1986). If we are facing a lion, this hyper-focus could save our lives but may be counterproductive otherwise. This human attentional bias was made famous by a 1999 study performed in a gymnasium where participants were simply asked to count the number of basketballs passed between players (Simons and Chabris, 1999). Dubbed, "Gorillas in our Midst", this study was a classic demonstration of inattentional blindness. During the basketball passing exercise, a person in a gorilla suit passed through the frame with most participants failing to notice.

Chronic anxiety and these cognitive biases are symptomatic of scarcity thinking; a survival mindset that pre-conditions us to focus our mental and physical energies on meeting the most pressing needs at the expense of other important goals (Tomm and Zhao, 2016; Mullainathan and Shafir, 2013). This can make us more effective in provisioning for basic survival needs but may create barriers to learning and other non-scarce activities. This is not a conscious process but one that competes in a bottom-up manner with the top-down directions for attentional focus. The scarcity mindset, as evidenced in this study, exhibits the following critical behaviors (Mullainathan and Shafir, 2013):

- Disproportionate thinking that captures the mind: creates attentional biases related to insufficiencies.

- Cognitive capacity (fluid intelligence) and executive control are reduced.

- Concurrent thought processes increase vulnerability to the bottom-up distractions and mind wandering

- Crisis orientation as short-term tunneling effects invokes further scarcity deficits.

Based on the scarcity/abundance research, finding ways of reducing scarcity thinking and inducing abundance thinking could be a critical strategy to address the negative consequences of scarcity mindset. Scarcity is an enigmatic concept where the very perception of scarcity or "perceived critical insufficiency" immediately appropriates scarce cognitive resources to address the need. (Mullainathan and Shafir, 2013). This self-propelling mindset can lead to increased and unnecessary demands on scarce resources.

## 2.2 Priming

The study by Mullainathan and Shafir (2013) illustrated that the most subtle priming techniques could have profound effects on our mental mindsets and cognitive capabilities. Priming occurs because the prime stimulus or action makes the content and subsequent cognitive processes more accessible, potentially influencing all stages of information processing: attention, comprehension, memory retrieval., inference, and response generation (Forster and Liberman, 2007; Wyer, 2008). The priming stimulus can be subliminal: brief and not easily detectable to the individual who is exposed, or supraliminal: detectable by the individual. Daniel Kahneman (2011) in his cognitive science research on system 1 (autonomic), and system 2 (conscious) thinking, describes priming as a process that affects our associative machine. Psychologists think of the associative machine as a vast network of memories in which each idea is linked to many others. Priming one idea or concept will activate many others, all happening concurrently and mostly at a subconscious level; like a ripple of waves on a pond, initiated by one idea that will activate through a small part of the vast network of associative memory (Kahneman, 2011). In our educational context, we define priming more generally to include any intentional action or stimulus, either cognitive (relating to cognition) or affective (relating to emotion), that strives to improve the student learning mindset. There are other salient features of priming that underlie the subtle and profound nature of these processes.

The ideomotor effect is an example of priming that initiates action from an idea. In a classic experiment by John Bargh and colleagues dubbed "The Florida Effect", one group of students was primed to create sentences with words related to the elderly. Words like Florida, balding, forgetful, gray or wrinkle were used to prime the concept of old age. The second part of the experiment involved walking to another room to complete a second task. The actual experiment was a deception. The intent of the study was to compare the time to walk to the second room, to the groups not exposed or primed by the elderly concepts. As predicted, the elderly primed groups walked much slower, exhibiting a significant behavioural priming effect (Bargh et al., 1996). This change in behaviour, resulting from primed ideas, is an ideomotor effect. The supraliminal nature of priming suggests that most effective priming is related to content that is activated above the conscious level (greater than 500 milliseconds) but below conscious awareness, not unlike the concept of "sleight of hand" used by magicians. A post questionnaire to the ideomotor aging study presented above asked students if they suspected that their test activities influenced their thoughts or actions in any way. Participants unanimously concluded that the study had no impact on their subsequent thoughts or actions, even though the slower walking results of the primed participants were significant (Bargh et al., 1996).

Conscious supraliminal activities that improve mood or mental states, demonstrate a positive impact on information organization and creativity. In a 1987 study (Isen et al., 1987), positive affect (elevated mood) was induced in study participants by viewing a few minutes of a comedy film or by receiving a small bag of candy. Another group received neutral stimuli while two more groups engaged in physical exercise meant to represent affective arousal. In performing two tasks that required creative ingenuity, the positive-affect groups, primed with candy or funny movies, saw improved performance while the control groups and exercise groups saw no performance increase. There is also evidence of the longer-term effects of frequent positive affect. In a 2005 UCR study "The Benefits of Frequent Positive Affect: Does Happiness Lead to Success?", the researchers demonstrated that happiness is associated with and precedes successful outcomes and may have a causal relationship with many of the desirable characteristics related to success (Lyubomirsky et al., 2005).

While positive affect does appear to mediate creativity, stimulating a creative process with randomness or a disruptive suggestion as with improvisation, may also offer similar effects to how we process information and the resulting creativity (Davies, 2019). Additionally, visualization of an experience, particularly from a 1st person perspective can actually improve performance of tasks that were not actually performed. The process of visual imagination is enough to generate real physical effects. For example, when people imagine writing, it will increase blood flow to those same brain areas activated with real writing activities. Similarly, imagining exercising can increase heart rate. This may seem like good news for procrastinators but both randomness and visualization can be effective methods for priming us to be better (Davies, 2019; Wong et al., 2009)

Psychologist Robert Zajonc created several studies demonstrating that mere repeated exposure of a word, concept, or image will enhance positive affect related to the initial stimuli and will even extend to similar or related concepts. He argues that this is a feature of most living organisms whose survival prospects are better served by avoiding novel stimulus (Zajonc, 1968; Monahan et al., 2000). As such, this evolutionary trait may create implicit biases that predispose us to stereotypical thinking that might be perceived as racist, xenophobic, and averse to that which is not familiar (Zajonc et al., 2001). This is also how norms are created. Consistent, repeated exposure to ideas or information induces cognitive ease (Kahneman, 2011), regardless of the moral value or veracity of the information. Eventually, the process will condition us to familiarity and enhanced positive affect related to these concepts. Celebrity culture is similarly based on this concept as McMaster University's Larry Jacoby and colleagues illustrated in his paper, "Becoming Famous Overnight" (Jacoby et al., 1989) with long term priming effects of name exposure. This human predilection for familiarity enables positive feelings for those concepts most exposed over other alternatives (Zajonc et al., 2001).

These and other studies show that various forms of priming related activities intended to improve mindset or academic performance can help reduce scarcity and improve abundance mindset. Emotionally preparing prior to the main activity (Gyllen et al, 2021; Mrazek et al, 2014), changing situational context (Lave and Wenger, 1991), allowing reflection (Di Stefano et al, 2016), motivating (Ryan et al, 2006), directing attention (Porter et al, 2010; McClelland et al; 2015), visualization (Davies, 2019, Wong et al, 2009), creativity (Isen et al, 1987; Bhagwatwar et al, 2018), and increased sense of inclusion (Peck et al, 2013, Banakou et al, 2016) are all viable priming intervention techniques.

## 2.3 Learning Theories and Mindsets

Behaviourism learning evolved from the work of early thinkers like Ivan Pavlov and B.F. Skinner, who introduced the concepts of classical and operant conditioning (Skinner 1971; Graham, 1982). Classical conditioning is a type of learning that is based on involuntary stimuli (e.g., scent, sound) where operant conditioning focuses on voluntary stimuli and learning because of choice and subsequent reward or punishment. Behaviourists offer great insights into aspects of priming that affect our cognitive processes. Classical conditioning theory is the basis for antecedent learning strategies where stimuli are primed or put in place prior to the learning experience to change behaviour. Antecedent strategies may include changing pre-learning or preparatory activities, rituals, or changing the environmental context of the learning environment (lighting, music, seating arrangements, décor) or any potential environmental stimuli that may affect the learner. While initially applied to deal with problem behaviours like autism spectrum disorders (ASD) (Conroy et al., 2005), these antecedent intervention strategies are now targeted more widely and used as a tool for improving academic performance (Kruger et al., 2016; Gyllen et al., 2021).

Cognitivism is more concerned with the process of learning rather than behaviour. Cognitive Psychology suggests that learning is more than a change in behaviour but is instead mediated by thinking that results in a change in understanding. The learner acquires knowledge and internal mental structures. Constructivism extends the previous theories by proposing a more subjective model of learning. With behavioural and cognitive theories, the world is real and external to the learner, whereas constructivism is a function of how the individual creates meaning from his or her own experiences (Ertmer and Newby, 2013). Experiential Learning Theory (ELT) and Situated Learning Theory (SLT) (Lave and Wenger, 1991) are extensions of the constructivist philosophy incorporating social context within the experience.

With Kolb's ELT, learning is cyclical and iterative: It begins with having a concrete experience, followed by a reflection of that experience, then the conceptualization of abstract concepts that incorporates the new insights from the experience with existing conceptual models, and finally, active experimentation of the lessons learned. The cycle continues to repeat as the learner's conceptual worldview is continually refined (Kolb, 1984). See below Figure I. This construct affords the opportunity for priming interventions within various points of the cycle and in multiple cyclical iterations. As such, ELT lends itself perfectly to a time priming methology.

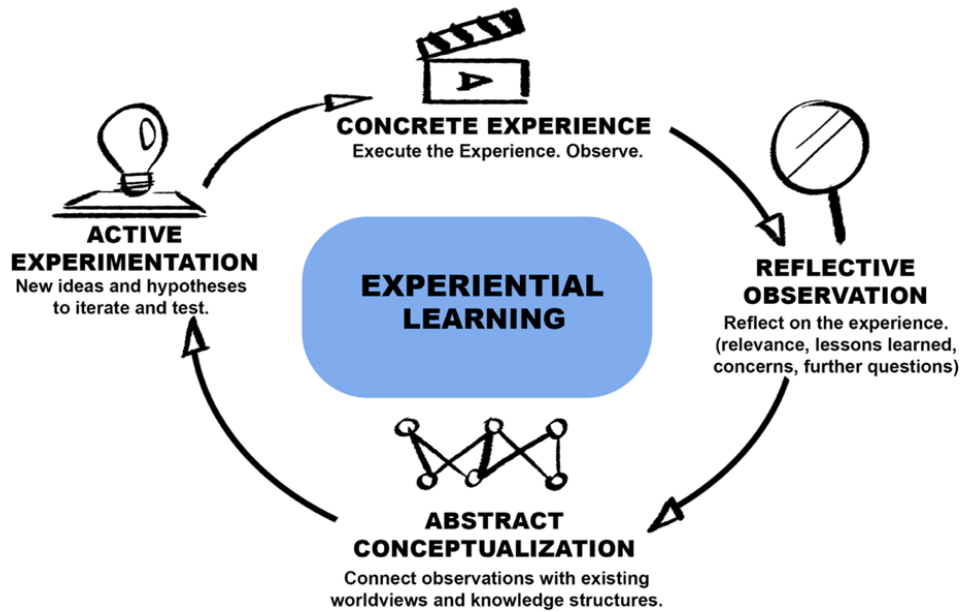

**Figure I. Experiential Learning Theory Cycle (Kolb, 1984)**

SLT advocates for learning within a community of practice and within a specific situation (Brown et al., 1989; Lave and Wenger, 1991). Securing meaningful situated learning scenarios, while desirable can also be a challenge, and made increasingly challenging through the COVID 19 period. Based on SLT needs, technology solutions that improve access and frequency to relevant situated environments and social contexts could create possibilities that would not be possible otherwise. These situations in turn could induce confidence for students through better understanding, increased motivation to try something new, and ultimately improve academic performance. For example, placing a student within an animation studio could alleviate emotional concerns like anxiety or low self-confidence with proper context and amplify the understanding of an animation related idea or concept.

Learning mindsets can have a considerable effect on academic performance. Growth Mindset (GM) research suggests that if children believed that their intelligence and academic success was malleable, organic, and in the process of growing (growth mindset), they progressed and achieved with greater success than those students who believed that their intelligence was "predetermined" (fixed mindset), and they were not in control of those variables (Dweck, 2008). Growth mindset is characteristic of abundance thinking whereas fixed mindsets are clearly a form of scarcity thinking. Dweck has furthered this research to indicate that disciplines like math, long considered the domain of the few, can be mastered simply by employing a growth mindset (Dweck, 2010). While growth mindsets are indicative of perceived control over academic achievement leading to more persistence and a willingness to embrace challenges, self-efficacy can be considered a related theory. Self-Efficacy Theory (SET) was defined by Albert Bandura as the belief in ones' ability to succeed at a particular task or challenge (Bandura, 1977). With SET, the willingness to embark on more significant challenges is indicative of a confidence level in one's ability to cope with the strain, incumbrances, and potential failure of the given challenge. Like growth mindset, high self-efficacy represents abundance thinking whereas low self-efficacy is indicative of a scarcity mindset.

**2.4 Technology Solutions**

In our technology research, we identify key technologies that have demonstrated the potential for improving student learning mindsets. VR and Video Games are already showing great promise and proving valuable in myriad application domains (Slater and Sanchez-vives, 2016; Fish et al., 2018). As such, we chose to focus on these. Virtual Reality (VR) is considered a subset of a more general concept, Extended Reality (XR); a term that includes VR, Augmented Reality (AR), and Mixed Reality (MR). VR is defined as images and sounds created by a computer that seem almost real to the user, that can affect them by using sensors (Oxford Dictionary, 2019). Central to the concept of VR is the quantifiable concept of immersion and the more subjective idea of "presence" or the feeling of "being there" (Slater, 2009). Based on these immersive possibilities and the ability to create a sense of presence, VR is an ideal environment to create scenarios that evoke different forms of conscious and unconscious awareness. Within a restricted domain VR can provide a blank canvas to allow us to create and control an environment and diminish distractions from other real-world influences (Slater and Sanchez-Vives, 2016).

VR-based tools have also been deployed in mindfulness studies. A 2019 study from Tokai University in Japan, tested game scores within a VR based virtual archery game following a 10-minute meditation using the *Muse* meditation headset (choosemuse.com). While there was no direct linkage to cognition, they observed improved game scores for all categories (Beginners + 275%, Intermediate, +107%, Experts +17). Further, after these activities were complete, all participants reported feeling recharged to continue their daily activities (Asati and Miyachi, 2019). Most research suggests that mindfulness does offer position psychological value, even with shorter sessions (Goleman and Davidson, 2017, Asati and Miyachi, 2019) but derives more benefits with prolonged practice. (Mrazek et al., 2014) In fact, there is evidence that meditators derive sustained neuropsychological effects like improved theta band activity in the right hippocampus, an area associated with memory processes (Lardone et al., 2018).

There is also growing evidence to suggest that VR may offer perceptual advantages that could not be acquired in the real-world. Participants embodied in an iconic VR avatar (Albert Einstein), performed better on cognitive tasks compared to the control normal body group suggesting that the design of avatars in the environment also present opportunities to elicit positive psychological effects that improve cognition (Banakou et al., 2018). Avatar embodiment techniques that offer participants the opportunity to inhabit custom personas of other races or cultures show great promise in reducing stereotypes, low self-perception, and negative academic performance associated with stereotype threats (Peck et al., 2013). Further, existing research indicates VR effects on reducing anxiety (Camara and Hicks, 2019), and improving academic performance and student experiences (Liu et al., 2020; Yu et al., 2018).

As a technology substrate, VR offers unique design and psychological affordances needed for realistic simulations, interactions, presence, and immersion (Shin, 2017; Steffen et al., 2019; Slater, 2009). VR provides the facility to design realistic, situational contexts: environments, artifacts, and avatars (Slater and Sanchez-Vives, 2016). The need to create custom physical environments to prime the learners would be very difficult and costly, if not impossible, without the use of VR. With VR, designers can create learning spaces that are more conducive to a specific learning objective. In a 2013 study, 3D Virtual Environments (3DVE) were used to test the effects of supraliminal priming on creativity (Bhagwatwar et al., 2013). The results showed that when teams created ideas in the primed 3DVE environments, they created more ideas, and the ideas were of higher quality than in the neutral., non-primed environments.

In popular culture, we associate games with being fun and escapist, which can seemingly diminish the potential positive impact of this technology. Today's games, particularly serious games, are far more complex and may contain positive pedagogies at various levels of play (Gamage and Ennis, 2018). Immersive worlds and simulation logic can allow educators to provide more realistic simulations that represent the learning context that they are attempting to convey. Games can be complex learning environments with the potential to mediate emotion and various cognitive processes.

Games can also have a direct impact on participants' mental states and performance. In a 2012 study, "Sparking Creativity: Improving Electronic Brainstorming with Individual Cognitive Priming", researchers demonstrated that supraliminal priming, a process where users are aware of the priming but not of the intent, did result in more creativity (Dennis et al., 2012). Members of brainstorming teams exposed to priming in web-based computer games generated significantly more ideas and of better quality than the neutral primed control group.

Games have been demonstrated to have therapeutic effects that can aid in reducing dependence on Selective Serotonin Reuptake Inhibitors (SSRI) anti-depressants. One of the most compelling examples of this was demonstrated in a 2018 study called Zombie's verses Anxiety: An Augmentation Study of Prescribed Video Game Play Compared to Medication in Reducing Anxiety Symptoms (Fish et al., 2018). In comparing two groups suffering from symptoms of anxiety, one group was prescribed one medication plus casual video game (CVG) play four times per week for 30-45 minutes for one month. The control group was prescribed a traditional two-medication regimen for controlling symptoms of S-Anxiety and T-Anxiety. (State and Trait Anxiety). Trait anxiety generally describes a personality characteristic rather than an ephemeral state of anxiety. State anxiety is anxiety related to an anticipated threat or fear. Results demonstrated that the CVG/one medication combination under a prescribed condition, significantly reduced S-Anxiety symptoms and moderately reduced T-Anxiety compared to the medication only, control scenario.

Gaming solutions have also been shown to improve working memory and consequently, cognitive capacity. The study results from this 2017 study, revealed that the extent of daily gaming activity was related to enhancement and speed of the most demanding level of working memory task (Moisala et al., 2017). A further study on the effect of anxiety levels on learning performance found that digital game-based learning was beneficial to high anxiety learners while there was a negligible effect for learners with low anxiety (Yang et al., 2018).

Finally, combining Artificial Intelligence (AI) with VR offers unique possibilities. While affective learning was once hailed as the panacea of human development (Picard et al., 2004), current VR implementations with high-fidelity immersive virtual spaces and customizable computational platforms are nearing ubiquity and offer the prospect of scalable priming interventions aimed at emotion regulation and academic optimization that can be personalized uniquely to each student and refined using AI algorithms for cognitive optimization (Picard, 2000) and/or facilitating antecedent requirements as with ASD (Conroy et al., 2005; Kern and Clemens, 2007).

## 3. Gap Analysis

The research suggests that to address the problem of student academic performance, we should focus our efforts on creating and testing technology solutions that ameliorate scarcity thinking (which includes anxiety) in favour of more positive abundance thinking to improve mental states and learning mindsets. The research also suggests that there are many other ways of reducing scarcity with priming methods. These include inducing positive affect, mental mindset conditioning, custom environment design, attentional control strategies, motivational messaging, and many others. The most proven and validated research studies appeared to be related to priming positive affect (Isen et al., 1987; Ashby and Isen, 1999), mental mindset conditioning (Mrazek et al., 2014; Hafenbrack et al., 2013; Goleman and Davidson, 2017), and improving environmental context (Bhagwatwar, 2013; Pena and Blackburn, 2013; Yu et al., 2018). In our first published study (Hawes and Arya, 2021), we investigated the efficacy of preparatory experience priming (PEP) through the use of gaming and meditation exercises that preceded the concrete experience. In both cases anxiety levels were reduced. We did observe academic performance increases in the post gaming activity but not in the meditation situation which was not totally unexpected as positive cognitive effects for meditation have been shown to be most effect over multiple interactions (Mrazek et al., 2015). Our second study (Hawes and Arya, 2022a) compared the priming effects within a situated learning environment. In this case, we observed positive effects of context-oriented priming (COP) in both IVR and DVR environments (Hawes and Arya, 2022b) that improved academic scores relative to students in non-primed environments.

What we lack however is more general framework and theoretical construct that accommodates new priming activities, a fertile environment like VR, and a timed, iterative experiential learning process, like ELT working in concert. As such, our learnings from our first two studies and subsequent reflection clearly indicates the need for a novel framework to further the longer-term prospects for this research.

## 4. Initial Research Studies

To address the problem of using digital technology (especially VR) to manage scarcity mindset in learners, we started with some basic questions:

- Can priming techniques transfer to a digital realm like VR?

- If transferable in general., what types of priming strategies and methods are most effective and how and when they should be applied? In particular:

    - Can VR-based preparatory priming (prior to learning) help with increasing academic performance?

    - Can VR-based context priming (throughout the learning experience) help with increasing academic performance?

To answer these research questions, we opted to break up the challenge into a two-phased approach. We then followed with a reflective approach to expand our theoretical model. Study 1 focused on the basic idea of priming positive mental mindsets to reduce scarcity thinking and its implementation using VR, while Study 2 sought to extend our understanding of priming strategies by testing context but also considering the effect of the immersion within the digital environment.

As such, we created an initial strategy framework that could provide the basis for a technology platform to accommodate our initial priming tests (PEP and COP) and future priming intervention theories. A concept diagram is presented in Figure II. The transformational process presented in the middle section of Figure 2 represents the ongoing priming research and proposed interventions used to transform scarcity thinking with limited cognitive capacity (red) to a more optimal learning state of abundance thinking, as indicated by the increased size of the rectangle (green) indicating increased cognitive bandwidth. The middle diagram in Figure 1 presents the first two priming strategies we used to create positive learning effects and facilitate improved academic performance. Initially, we focused on preparatory (early) interventions and context-oriented interventions,

- PEP - Pre-Learning or Preparatory Experience Prime
- COP – Context-Oriented Prime

Preparatory Experience Priming (PEP) is executed prior to the educational experience or activity (Dennis et al., 2013; Gyllen et al., 2021) whereas Context-Oriented Priming (COP) strives to create a realistic environmental context based on situated learning theories (Brown et al., 1989; Lave and Wenger, 1991) during the concrete experience. An example of a PEP would be a short game, nature walk, or funny video. A COP could be a custom-designed creative environment that is aligned with the learning content and allows the student to feel more creative, like learning animation storytelling within a virtual animation studio. The common details of both initial studies are provided in (Hawes and Arya, 2021; Hawes and Arya, 2022a).

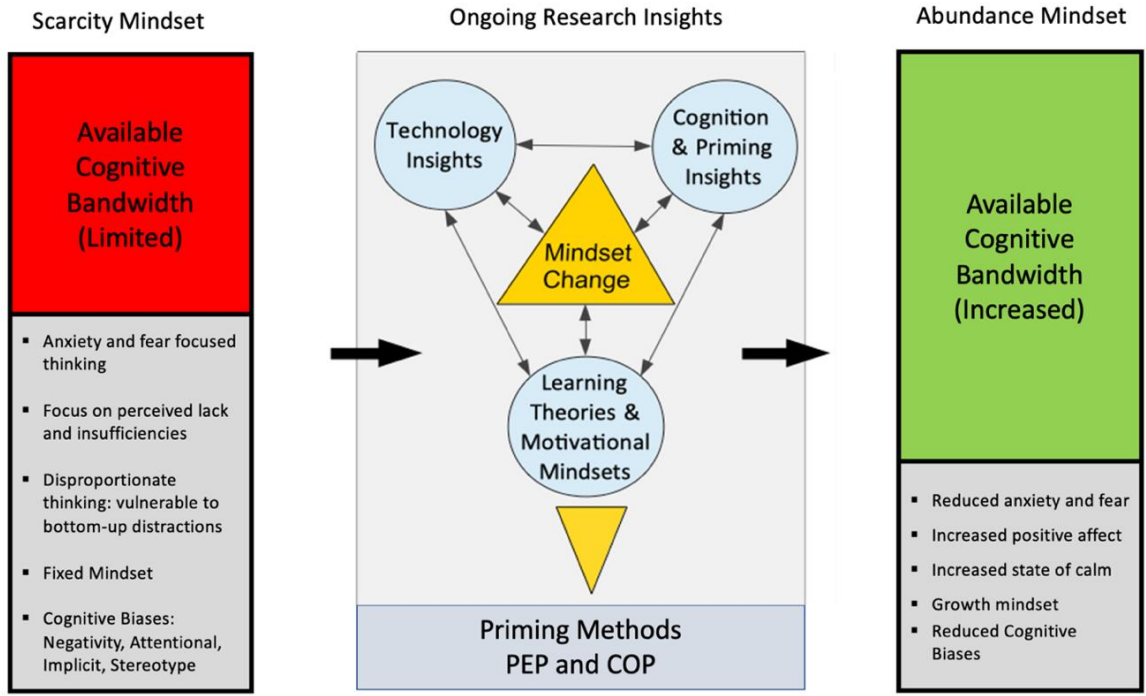

**Figure II. Study 1 – Gaming and Meditation Preparatory Experience Priming**

**4.1 Preparatory Experience Prime (PEP)**

**4.1.1 Overview**

Our PEP study (Hawes and Arya, 2021) was focused on the efficacy of using VR for preparatory priming. Given that Meditation and Games have been shown to have potential in other digital and non-digital contexts to reduce anxiety and increase cognitive bandwidth (not necessarily correlated), we hoped to determine whether these priming methods do work within fully immersive VR environments and their relative efficacy. We were also curious to gain more insight into the emotional and performance aspects of these priming methods. We defined our research hypotheses as follows.

**H1a.** VR Meditation can reduce anxiety levels (scarcity mindset).

**H1b.** VR Meditation can increase cognitive bandwidth.

**H2a.** VR Gaming can reduce anxiety levels (scarcity mindset).

**H2b.** VR Gaming can increase cognitive bandwidth.

**H3a.** VR Gaming will be more effective than VR meditation for reducing anxiety.

**H3b.** VR Gaming will be more effective than VR meditation for increasing cognitive bandwidth.

We designed a mixed method experiment with two independent variables: (A) priming method (game vs. meditation) and (B) priming condition (prime vs. no prime). The 2x2 design resulted in a mixed-method (between and within-subjects) research study. The dependent variables (measurements) for each of the four cases were anxiety levels as measured by a shortened version of the State-Trait Anxiety Inventory (STAI) (Marteau et al., 1992) and cognitive bandwidth improvements as measured by the University of California Matrix Reasoning Test (UCMRT) (Pahor et al., 2019). Participants were divided into two groups based on variable A and were asked to complete both options of variable B (prime and no prime conditions for both game group and meditation group participants). This allowed us to not only compare the effects of two priming methods but also compare this effect with the no-prime condition for the same participants. More specific details of the study setup can be found in our paper (Hawes and Arya, 2021).

We chose to use VR gaming as a tool to prime positive preparatory experiences (PEP). This choice is based on the demonstrated potential of video games to improve the player's emotional state (Fish et al., 2018; Yang et al., 2018). One such game that demonstrated that appeared to have priming possibilities was *Beat Saber*. The *Beat Saber* game provided a fully immersive flow game experience that had been described as both invigorating and fun. Further, the feelings of immersion and presence were well executed on the Oculus Quest. A visual of the game is presented in Figure III. The process was very simple and lasted 2 minutes and 35 seconds. The participants use their *beat saber*s to slice oncoming blocks and circles in an optimal manner while energetic music plays in the background. We requested users play the demo round twice for a total of 5 minutes and 10 seconds, then commence to the next activity.

The Meditation app was a custom-designed guided meditation application inspired and informed by the CALM app and recent research on mindfulness app evaluations (calm.com; Roquet and Sas, 2018; Goleman and Davidson., 2017; Huberty et al., 2019). As the participant enters the meditation space on a lofty mountaintop above a large body of water, they are met with beautiful nature sounds (waves, wind, and birds chirping), and after 15 seconds, a voice-guided meditation commences. The meditation process lasted for 5 minutes, at which point participants were asked to take off the headset to begin the next step. While we considered testing for a longer period of time, we chose a 5-minute interval because we preferred a shorter experience, comparable in time with the gaming experience (5 mins, 10 seconds)  and an activity that a student might be able to do prior to class. Further, research suggested that 5-minute meditation sessions can be effective (Krygier et al., 2013) and viable prior to classroom sessions (Ishikawa et al., 2020). Recent successful meditation products have also demonstrated success with 5-minute sessions. (choosemuse.com). An image of the custom Meditation app is presented in Figure IV.

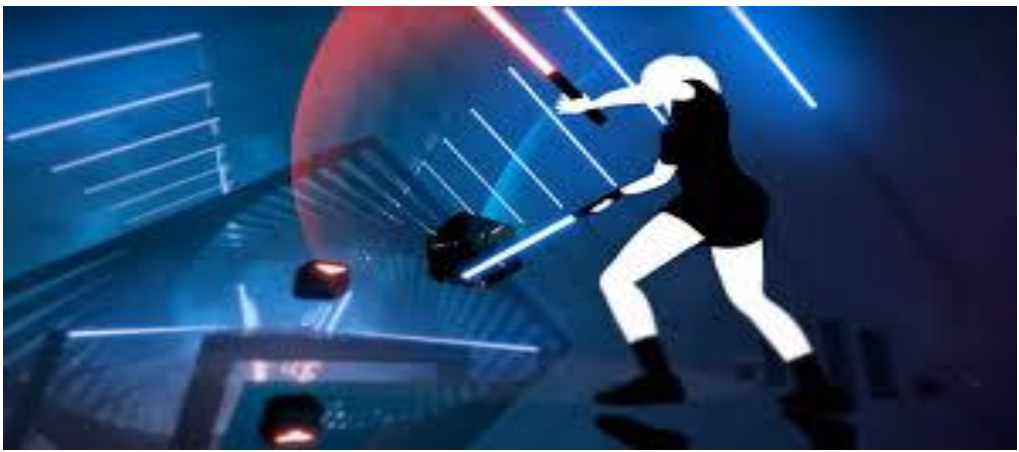

**Figure III. The Beat Saber Game**

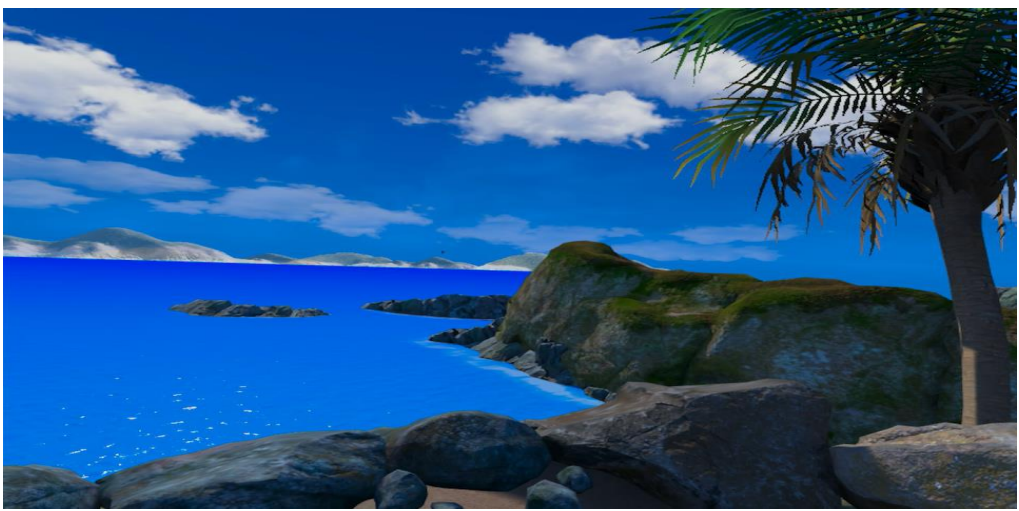

**Figure IV.  The VR Meditation App**

### 4.1.2 Results

This experiment design resulted in four sets of measurements and required a mixed model, so we ran a 2-factor ANOVA with one repeated measure for both anxiety and cognitive score data to see if there was any significant difference in between-subjects and within subjects-data. These measurements were done for both cognitive bandwidth (objective) and anxiety (subjective). They were all continuous data and had a normal distribution. This allowed us to use parametric methods such as ANOVA. The ANOVA results suggested that priming was effective for anxiety, with both methods performing with an insignificant difference from one another. On the other hand, for the cognitive tests, while priming was effective overall, there was a difference between the two methods. To support ANOVA results for our key area of study (academic performance), separate post-hoc paired t-test comparisons for meditation and gaming cognitive test scores demonstrated significance indicating a positive improvement in test scores due to VR game priming (P=.0008), but no significance was observed indicating no positive improvement in test scores due to VR meditation priming (P=.34). A similar test for anxiety showed significance for both games and meditation, with P values of .02 and .01, respectively. A further T-test comparing deltas of prime vs. no-prime between VR Meditation and Gaming for anxiety reduction yielded no significant difference (P=.41). Overall, the T-Test did not provide us with any new and helpful insight. We used a two-tail test in all these cases to allow for detecting any significant differences in either direction. To distribute the practice effects, we did interweave the baseline testing of the UCMRT by placing the no prime conditions alternately before and after the priming experience.

### 4.1.3 Discussions and Limitations

All hypotheses were supported except for H1b (meditation improves cognitive bandwidth) and H3a (games are more effective than meditation for anxiety). While the anxiety assessment was subjective, participants felt better after the priming exercises (both meditation and gaming) than before. Existing research suggests that positive affect and mental calm can effectively reduce anxiety. While this may result in better cognitive performance, it was evidently not causal in this one-time meditation scenario. Upon reflection, the VR meditation activities' limited effect on cognitive bandwidth was unsurprising. Like sports or crafting, meditation is a long-term skill that yields more benefits over time as participants improve and better understand how to apply the process. The academic trials that did observe improved cognition demonstrated benefits over a longer time. For example, GRE reading comprehension scores were increased after a 2-week mindfulness training course (Mrazek et al., 2013). On the other hand, the gaming activities were more ephemeral and generated an immediate elevated state. Further, as in the Isen studies that compared the act of receiving candy, funny videos, exercise, and neutral movies (Isen et al., 1987), exercise did not generate similar positive creativity effects. As such, we theorize that the gaming effect was most likely attributable to increased positive affect. From this research, the promise of immersive within VR technology did clearly improve emotional states in both the game and meditation states and as such, it could serve us well to create a technology construct that would allow us to iterate and reconstruct those ephemeral emotions on more sustainable basis not unlike how Virtual Reality Exposure Therapy iterates exposure to fear, phobias, and areas of discomfort to habituate and desensitize. Further, an iterative technology construct would suit a similar educational process like ELT.

## 4.2  Context-Oriented Prime (COP)

### 4.2.1  Overview

Our COP study (Hawes and Arya, 2022a) was focused on investigating the effect of VR and learning context. The study, a between-subjects test with one independent variable (Priming context), explored the effectiveness of context-oriented priming (COP) within the ELT cycle. The study used three dependent variables (academic score, user experience, and affective improvement). Academic score was our main subject and represented the learning effect of the context. It was measured through a short test after the course. We measured the subjective user experience (UX) satisfaction through a survey to investigate any possible effect. Anxiety and positive affect were also measured using a pre-course and post-course short-form State Trait Anxiety Inventory (STAI) (Marteau et al., 1992) and Positive Negative Affect Schedule (PANAS) (Watson et al., 1988) surveys, respectively. The STAI measured changes in self-reported anxiety levels while the PANAS changes in both positive and negative affect. Based on anticipated priming and situational learning effects, we investigated the following hypotheses:

**H1:** (a) Prime conditions (both) will improve academic performance over the no prime condition. (b) There will be no significant difference between the two priming contexts.

**H2:** (a) Prime conditions (both) will improve subjective assessments of UX results compared to the no prime scenario. (b) There will be no significant difference between the two priming conditions.

**H3:** All three conditions will improve affective state, as it has been shown in our previous work that enjoyable VR experiences can decrease anxiety.

### 4.2.2 Results

We had a 1x3 experiment design with one independent variable (Priming condition – prime vs. no prime) and three dependant variables (academic performance, UX satisfaction, and affect). After reviewing the outlier data, we opted to eliminate the highest value from each of the three conditions to avoid skewing results based on potentially contaminated data. These high values occurred for academic scores and were more than twice the average. They came from participants who were very familiar with the subject. Hence our sample size was reduced to 16 participant inputs per condition. With an equal number of participants (16), we first conducted a single factor ANOVA test for academic scores and data samples. After confirming the significance of mean departure within the conditions (p=.0099), we performed a post-hoc t-test with 2 samples, assuming equal variance. We observed significance in both priming conditions compared to the no prime condition, indicating that participant scores in both prime conditions were significantly higher than in the no prime condition: (prime (artifacts) vs. no prime (P=.0023) and prime (animation studio) vs. no prime (P=.0103)). With T stats of 3.05 and 2.44, respectively (>2.0), we were confident with the results. As hypothesized, participant test scores between the two prime conditions were comparable as no significance was observed between these two conditions (P=.258). Final comparison statistics for both priming conditions (VR theatre classroom and Animation studio) compared to no prime conditions did not observe significance, in either case. Similarly, both PANAS and STAI did not observe significance in either case comparing the primes to the no prime condition.

### 4.2.3 Discussions and Limitations

Our hypotheses were designed to test the academic effects of our primed VR environments compared to our no prime condition. In reviewing the first hypothesis, both prime conditions observed significant improvement over the no prime condition indicating that primed VR environments did have a significant positive effect on academic scores. There was no significant difference in test score performance between the two priming contexts. Hence, H1a and H1b were both supported. In reviewing the second hypothesis (UX), neither prime condition demonstrated significance over the no prime condition. Hence, H2a was not supported. The prime conditions observed no significance between themselves with respect to UX results. Hence, H2b is supported. While we considered the possibility that artifacts and situated learning environments might improve affect, resulting in a better-perceived experience, this was not the case. In all conditions, affect did improve, but there was no significant difference between conditions. This was not surprising based on our understanding of the unconscious nature of priming and consistent with the research (Bargh, et al., 1996; North et al., 1999).

There was a great deal of relevant feedback provided through the UX study that will inform future design. Image resolution appeared to be a distraction for some, likely decreasing the presence effect for those so affected. Striving to eliminate issues of blurriness, aliasing, or any image resolution should be a major focus in future environment and character designs for VR spaces. Part of the challenge here was that one "size does not fit all" with VR headsets. Some participants would have benefitted from optical lenses that were specifically made for visually challenged participants. In a non-COVID study, where participants would have had immediate feedback from researchers, some of the visual issues could have been mitigated on the spot. Interestingly, participants seemed to favour the familiar as several comments lauded the similarity to university lecture halls and the overall university learning experience. This insight suggests that students, as per Robert Zajonc's Mere Exposure theory (Zajonc, 1968), people favour the familiar and this fact should also be regarded in future designs.

Finally, as accessibility is critical and not everyone can use a VR HMD, assessing the effectiveness of priming in non-immersive VR spaces could inform the design of new or varying priming strategies. To address this, we did conduct a supplementary test to determine if immersion was a factor in the efficacy of the priming effects. By comparing the prime based on the best result from Study 2 with an identical Desktop VR prime, we discovered that immersion did not appear to be a factor. The results for both prime and no prime conditions did not observe a significant result. Hence, we concluded that for this scenario, context priming was equally effective in IVR and DVR environments.

5. **Cyclical Priming Methodology**

Our literature research focused on how students learn, how they think, and how they use technology and media. We discovered that both IVR and DVR environments afford the ability to create scalable, realistic, digital environments, and simulations that could affect perception and induce desired emotional mindsets (Hawes and Arya, 2021; Hawes and Arya, 2022a). Further, these scalable mindset priming interventions can be personalized to specific learners' needs, no different than students with ASD or related learning challenges that require very specific antecedent conditions in their learning environments (Conroy et al., 2005; Kern and Clemens, 2007). What is more, the profound nature of supraliminal priming offers an effective and non-invasive method of positive intervention. Hence our VR-based priming approach can offer an unobtrusive implementation to improve student mindsets.

In our previous studies we tested preparatory priming and context priming validating the effectiveness of the priming effects in both cases. While traditional research suggested that these desired performance effects would prevail, we were unsure of the transferability to these digital environments (both immersive and non-immersive) and within the specific situated learning contexts. Hence these findings were novel. In the PEP case, immersive gaming proved an effective psychological elixir to reduce anxiety and increase cognitive bandwidth prior to the learning experience. The nature of the Beat Saber VR video game ([www.beatsaber.com](www.beatsaber.com), 2018) used in our first study was flow-based (Csikszentmihalyi, 2000) and repetitive in nature as participants used their beat sabers to "chop blocks" to music, both directionally and based on colour. The iterative aspect of the game appeared to have a positive emotional effect as all participants were effusive in their feedback suggesting that priming activities would most likely benefit from repetition and iteration. As such, procedural simulation technology like VR appeared to be a good technology fit to build upon.

The COP study used situated learning environments and subject matter artifacts to prime the desired learning effects in both IVR and DVR that would be expected within a real-life situated learning environment. While self-reported tests related to anxiety and positive affect did not observe increases compared to no prime conditions in study 2 we observed a difference in self-reported motivation and increased self-confidence to embark on further challenges related to making animated movies compared to no prime conditions. While we did observe a similar effect in study 2 (COP IVR), it was only observed in the artifact prime condition. Nonetheless, this suggests that increased motivation and improved self-confidence might be the mediating emotional variable in situated VR learning environments. We also validated that VR could become a powerful tool to deploy for experiential learning purposes. Particularly during COVID were coveted cooperative university programs like the University of Waterloo (uwaterloo.ca, 2022) simply were not accessible to students or businesses leaving a significant gap in the transfer of knowledge and confidence into the workplace. Hence, the combination of VR's powerful affordances of simulation, presence, iteration, and immersion offer an alternative solution. As such, in Study 2, we were able to create and test the situated learning effects of custom designed and primed environments that would be prohibitively expensive to build, modify and deploy regularly.

In summary, from our multidisciplinary literature review and two studies we learned that:

1. Affective elements of real-world learning experiences do transfer to the digital domain (Hawes and Arya, 2021)
2. VR presence, and immersion offer positive emotional effects to students (Steffen et al., 2019; Slater, 2009).
3. Mere repetitive exposure to positive (or non-negative) priming stimuli can reduce scarcity in favour of abundance thinking improving student mindsets and academic performance (Zajonc, 1968; Lamb et al., 2019; Yeager et al., 2016)
4. VR tools offer powerful affordances to create situated learning environments (Slater and Vive-Sanchez, Lave and Wenger, 1991; Hawes and Arya, 2022a), otherwise not be economically viable or scalable.
5. The ELT Cycle is a suitable substrate to apply key repetitive requirements of our priming interventions.
6. A novel theoretical construct should incorporate cyclical iteration, virtual affordances, and innovative priming strategies that improve student mindsets and academic performance.

Our PEP and COP priming interventions were proven to be effective within our VR substrate. As such we plan to expand with additional priming intervention strategies for future research.

**6. Combined Insights for an Improved Model**

Given the insights (above), we propose a comprehensive priming model called Cyclical Priming Methodology (CPM) that accommodates the following emerging requirements/features, as shown in Figure V below:

- Expand our Scarcity and Abundance research model to include more proactive, repetitive strategies to facilitate systematic mindset transformation.

- Apply the priming interventions throughout the ELT process. For example, due to cyclical nature of ELT, it is important to offer motivational priming throughout the learning activities and reflective priming after each element or cycle of learning.

- To offer flexibility and extensibility, provide the possibility of "plugging in" other forms of priming into CPM. For further research, we propose including interventions such as inclusiveness, creativity, attentional controls, and visualization.

- Implement using technological solutions such as VR.

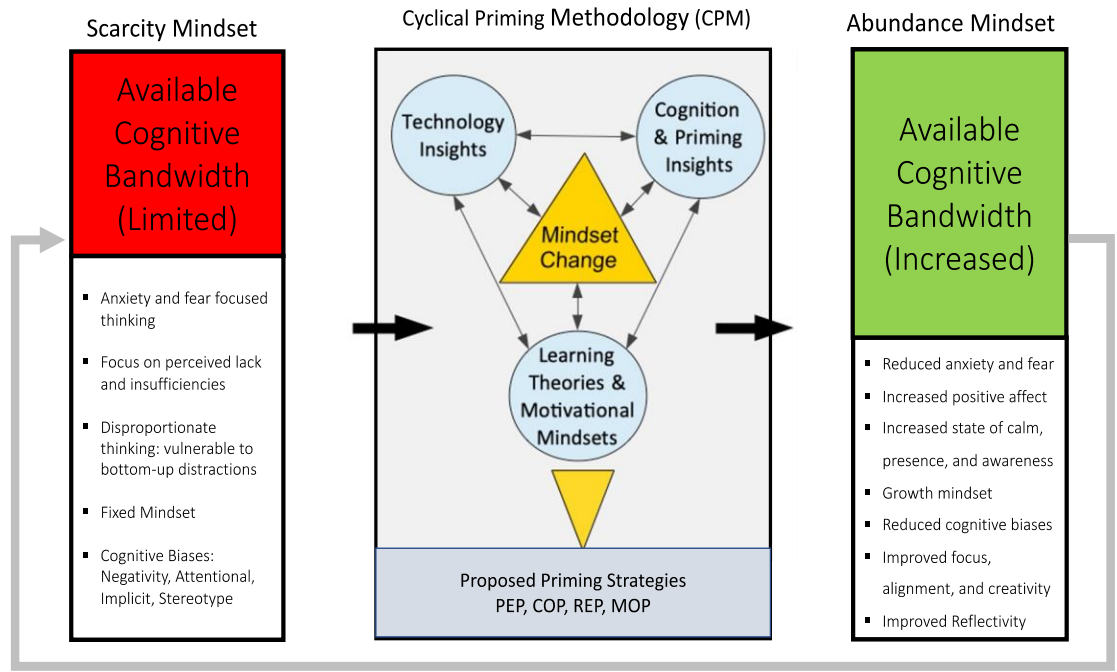

Figure V.  Cyclical Priming Methodology – CPM

While traditional priming methods mostly focus on preparatory (early) interventions and context-oriented interventions, the CPM proposes adding additional priming methods to the priming process. Figure 5 (above) presents two additional priming methods that could be explored next.  A post-learning or reflective experience prime (REP) might be a prompt that asks the student to consider the relevance to a real-world situation that they may have experienced. The reflective activity could help the student understand the relationship between the learning content and real-world challenges. A motivation oriented prime (MOP) would be delivered within the context of the learning experience, perhaps on the periphery of a content screen as simple as a "positive image" or a motivational suggestion delivered within the context of the learning modules. Example MOP images are presented below in Figure VI.

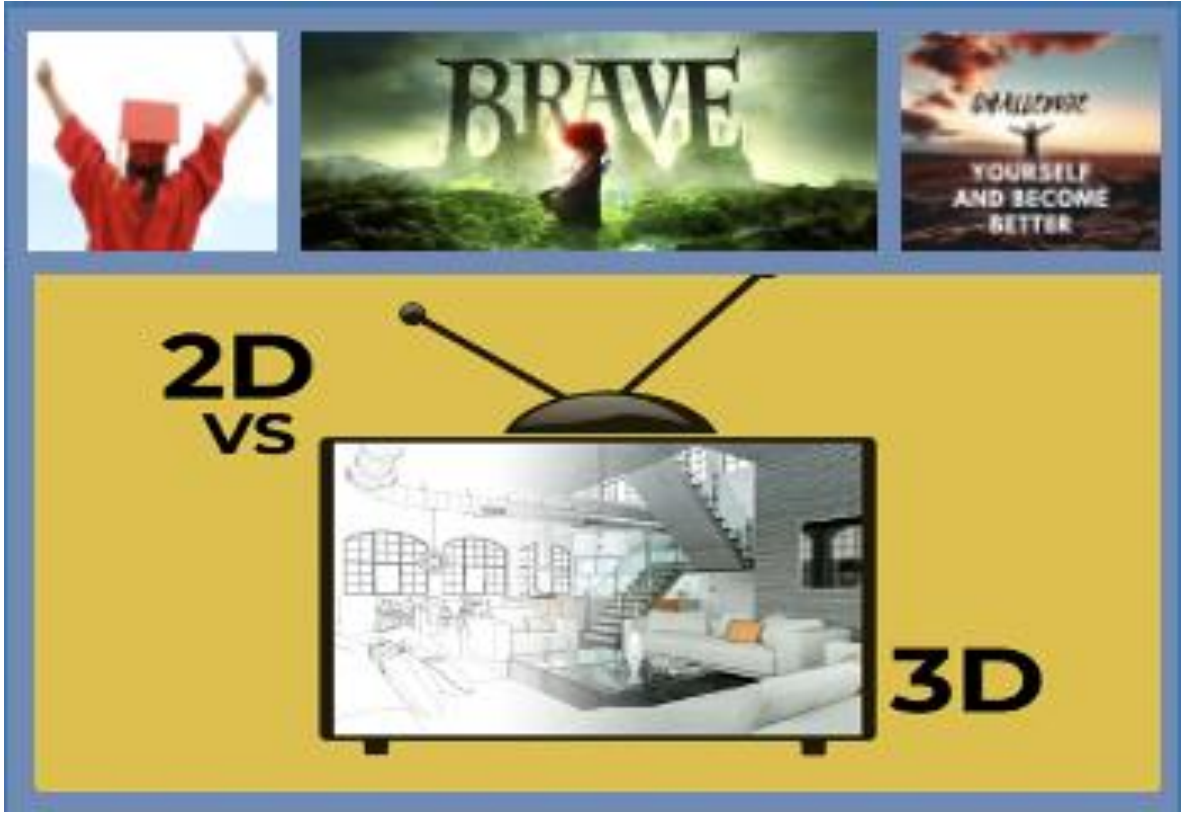

**Figure VI. Example Motivational Images on Screen**

### 6.1 Virtual Reality Educational Priming (VREP)

To accommodate the technology needs of CPM, we proposed a virtual reality experience priming (VREP) technology model and substrate. VREP is a specialized educational model: a specific implementation of CPM that uses iterative, simulative capabilities, and virtual worlds of VR (Freeman, 2018; Hoffman et al., 2000) to deliver relevant priming interventions that would otherwise not be viable. Our model builds off an existing frame of reference, Virtual Reality Exposure Therapy (VRET) (Carl et al., 2019). VREP could be perceived as an educational., non-therapeutic version of VRET. While VRET relates to cognitive based therapy (See Figure VII), VREP is based on the concept of cognitive bias modification.

VRET utilizes habituation techniques; an iterative process of exposure to fear stimulus (Carl et al., 2019; Thompson, 2009; Wagner, 1979). Once that level of fear stimulus is diminished, the stimulus will be elevated to a higher level and re-iterated until the fear or phobia is fully extinct.

VREP similarly uses iteration and repeated exposure techniques but for differing objectives: educational vs. therapeutic. Where VRET focuses on a therapeutic context using repetitive exposures in the absence of negative consequences to reduce fear, VREP focuses on an educational context using positive emotional exposures to achieve optimal mental states for learning or academic performance. Both exploit the power of repetition, exposure, and immersive VR.

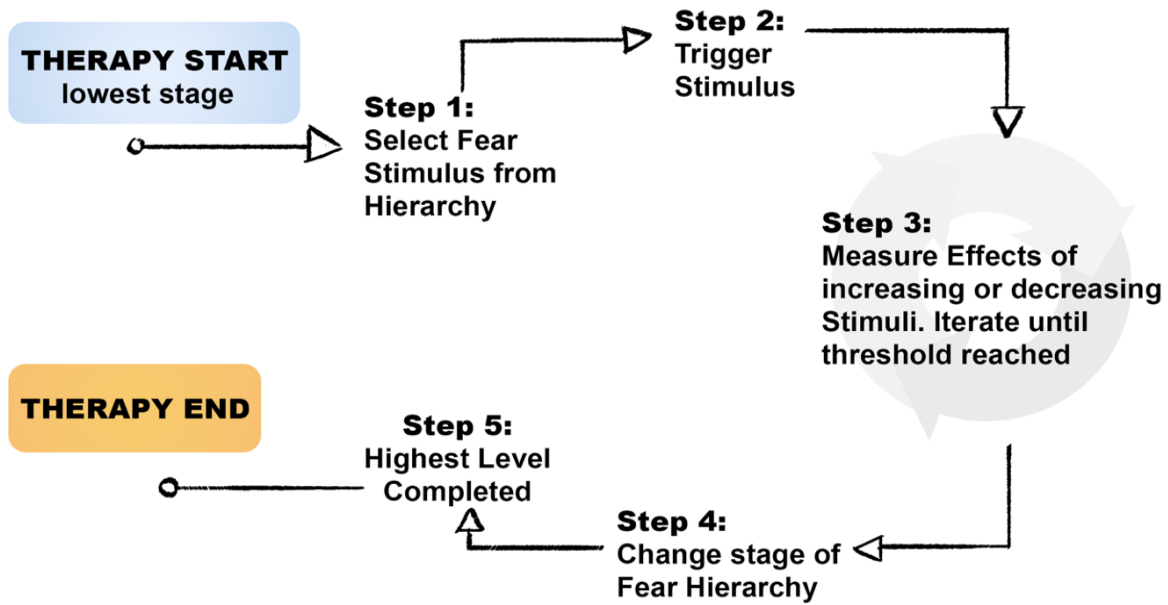

**Figure VII. VRET Process: A proven model (Carl et al., 2019)**

While VRET may require many iterations to create the desired effect, the VREP process is more ephemeral and can have an immediate impact on emotion and cognition. ELT views learning as an iterative process where students continually reconstruct experiences by learning and relearning. Even if cognitive priming strategies ignite only small changes in mental state, if the timing is right, these mental jolts may be enough to improve engagement for the duration of the learning process. These strategic priming interventions, executed before, after, and within the learning experience must feel seamless and natural to the student. A conceptual visualization of VREP in the context of a VR world with some examples (PEP, REP, MOP, COP) is presented in Figure VIII.

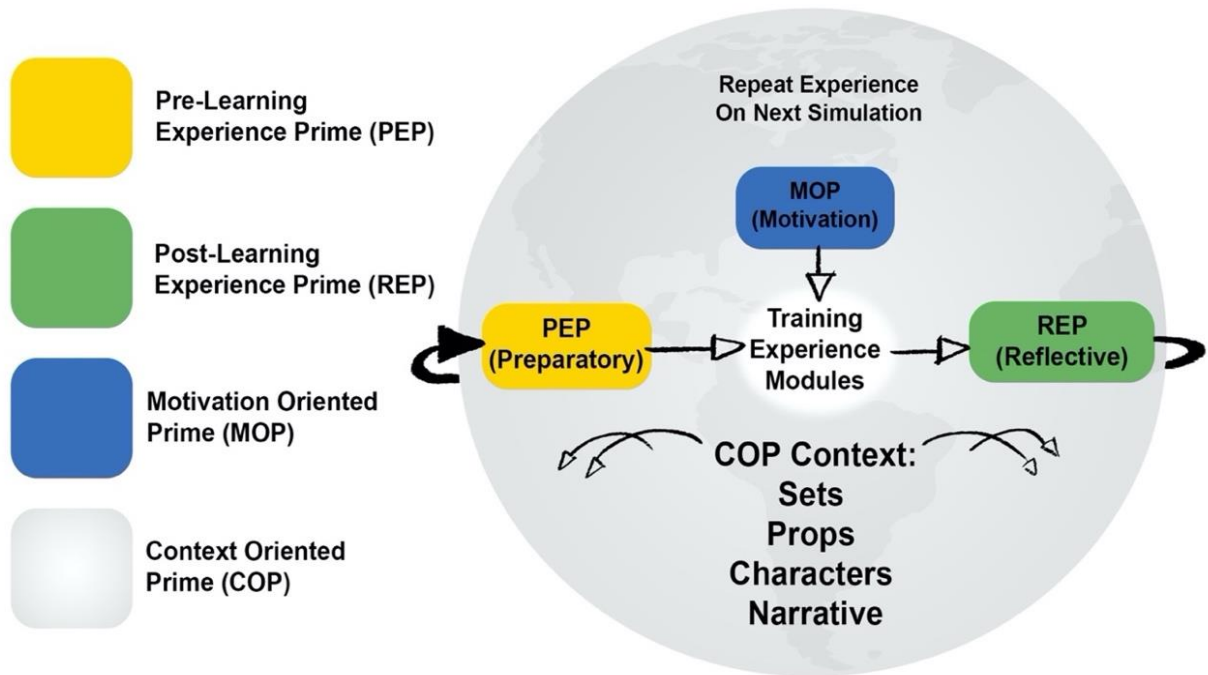

**Figure VIII. VREP: Example priming Interventions within an immersive VR World**

## 6.2 Future Priming Methods to Consider

While our initial research initially considered four types of prime (PEP, COP, REP, MOP) to be applied within the ELT cycle context, there are many other possible priming methods that could be considered to proactively improve and optimize elements of the learning experience (creativity, inclusivity, attentional focus, and positive affect).

For example, while positive affect has been shown to induce creativity, stimulating a creative process with randomness or a disruptive suggestion (improvisation) may improve how we process information and the resulting creativity. Additionally, visualization of an experience, particularly from a $1^{st}$ person perspective can improve performance of tasks that were not actually performed. The process of visual imagination is enough to generate real physical effects. (Davies, 2019). For example, when people imagine writing, it will increase blood flow to those same brain areas activated with real writing activities. Similarly, imagining exercising can increase heart rate. This may seem like good news for procrastinators but both randomness and visualization can be effective methods for priming us to be better. We may refer to the related forms of possible priming as creative improvisational priming (CIP) and visual imagination priming (VIP).

Robert Zajonc, (Zajonc, 1968) argues that avoidance of others, that manifests in myriad ways (racism, sexism, ageism, etc) is a feature of most living organisms whose survival prospects are better served by avoiding novel stimulus (Zajonc, 1968; Monahan et al., 2000). As such, these implicit biases that predispose us to these stereotypes, and negative norms could be addressed with mere repeated exposure to ideas or information that induce cognitive ease (Kahneman, 2011) and improved acceptance of the "other". As such, perhaps a unique context priming intervention that called prosocial inclusiveness priming (PIP). Finally, direct attentional priming (DAP) is another possible intervention that helps with distraction and draws attention to a subject to enhance real-time engagement.

As discussed above, additional CPM priming interventions could be researched, validated, and applied within the ELT Cycle as in PEP and COP (Figure IX and Table I).

**Table I. CPM Priming Intervention Summary**

| Cognitive & Affective Priming Strategies for Improving Student Mindset and Academic Performance applied within ELT ||||||
|---|---|---|---|---|
| **Priming Categories** | **Example Priming Activities** | **Recommended ELT Timing** | **Prime Type** | **Key Focus** |
| **Preparatory Priming (PEP)** | Games, Meditation/Mindfulness, Movement/Exercise, Stories, Mysteries, Humour, Gifts | Pre-Concrete | Real-time Emotion Regulation | Affective |
| **Context Oriented Prime (COP)** | Situated Learning Environments, Characters, Artifacts, Sounds/Music, Light, Embodiments | Concrete | Supraliminal/Subliminal Priming | Cognitive/Affective |
| **Prosocial Inclusiveness Priming (PIP)** | Design of Inclusive Environments, Characters, Social/Cultural Artifacts, Embodiment | Concrete, Pre-Reflective, Reflective | Supraliminal/Subliminal Priming | Affective/Cognitive |
| **Motivation Oriented Priming (MOP)** | Associative Priming from Positive Words, Imagery, Sounds/Music, Challenges, Mystery | Concrete, Pre-Reflective | Direct Calls to Action | Affective/Cognitive |
| **Reflective Experience Priming (REP)** | Questions related to Learning, Theories, Real World Implications or other considerations, | Pre-Reflective, Reflective | Direct Calls to Action | Cognitive/Affective |
| **Direct Attentional Priming (DAP)** | Distractions or calls to attention to enhance real-time engagement or awareness of critical norms | Concrete | Direct Calls to Action | Cognitive |
| **Visual Imagination Priming (VIP)** | Visualizing and Imagining an activity being performed and the intended result | Pre-Abstract, Pre-Experimentation, Pre-Concrete | Direct Calls to Action | Cognitive/Affective |
| **Creative Improvisational Priming (CIP)** | Thought experiments, planned or random changes that disrupt the process requiring improvisation | Pre-Abstract, Abstract, Pre-Experimentation, Experimentation, Pre-Concrete | Direct Calls to Action | Cognitive |

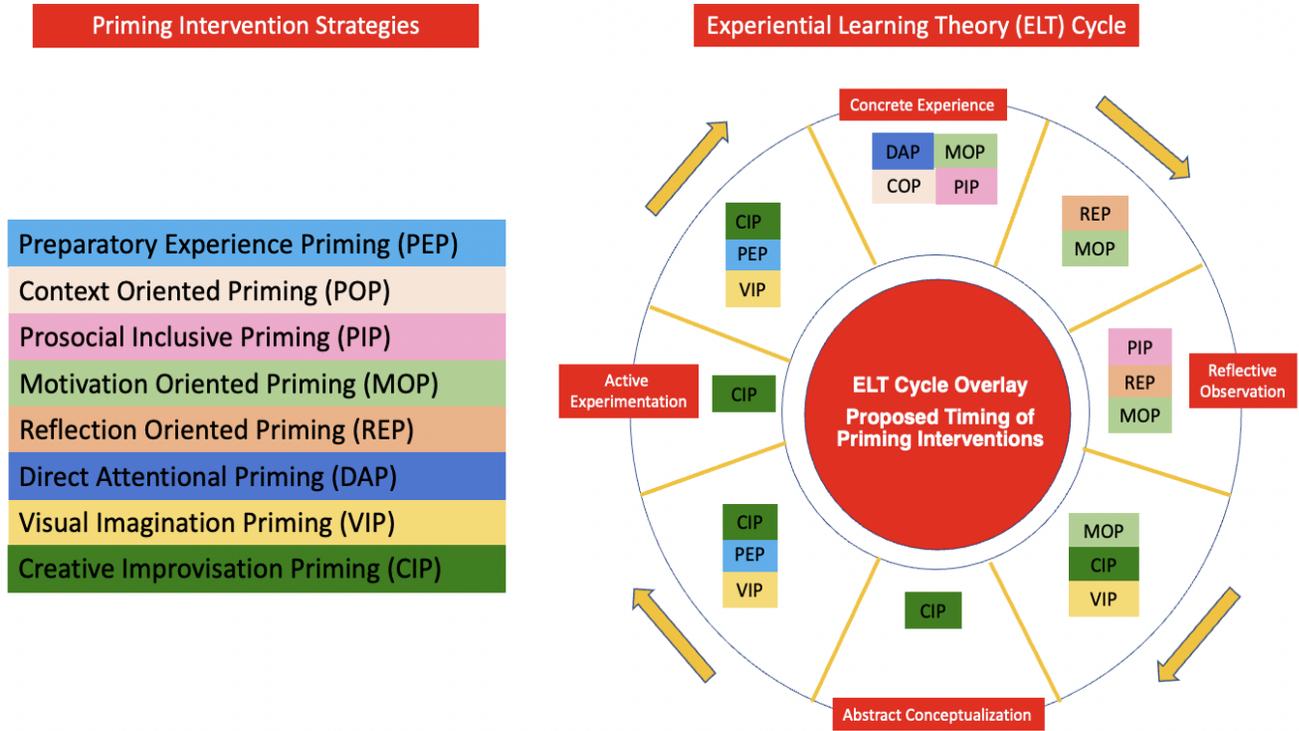

**Figure IX. ELT Overlay**

## 7. Conclusion

Our research was motivated to find technology solutions to minimize the impact of scarcity thinking (anxiety, negative biases, stereotypes, and negative mindsets): a counterproductive mindset that diverts precious cognitive resources needed for important activities like learning. Through the lens of *Scarcity Mindset*, we conducted a multi-disciplinary literature analysis of innovative ideas in cognitive science, learning theories and mindsets, and technology approaches suited to address the challenges of scarcity thinking. We identified priming strategies that help transition students to a more positive mental mindset. Based on this initial priming research model, we reviewed two past studies based on preparatory experience priming (PEP for gaming and meditation) and context-oriented priming (COP for situated learning environments).

In our post study reflection, we proposed a comprehensive theoretical priming framework called Cyclical Priming Methodology to be applied within the ELT Cycle and a VR implementation of CPM called VREP that proposes priming interventions throughout this experiential learning cycle. We discussed various forms of priming that can be supported by CPM/VREP. Further research is required to investigate the effectiveness of the model and suggested interventions and to explore other forms of priming that can be added.

Additional research questions for future research are:

- **Who** are the most sensitive or suitable to the priming interventions?

- **What** subject matter content is most sensitive and favourable to these priming techniques?

- **Where** are these priming techniques best applied (e.g., nature environments, creative environments, historical settings, specific geographies, or landmarks that evoke emotional responses)?

- **When** is "prime time"? How would we apply these various methods within the ELT cycle? Can we have priming during and after the main task? How frequently should priming be applied?

- Investigating the effect of various types of short games and gameplay features as priming methods (e.g., Cognition, Problem-solving or physical movement in 3D)